\documentclass[12pt,thmsa]{article}
\usepackage{amsfonts}
\usepackage{amsmath}
\usepackage{makeidx}

\begin{document}

\title{Noncommutative Geometry and Symplectic Field Theory }
\author{R. G. G. Amorim$^{a}$, M. C. B. Fernandes$^{a}$, F. C. Khanna$^{b,c}$,
\and A. E. Santana$%
^{a}${\thanks{Corresponding Author. E-mail: asantana@fis.unb.br}}, 
J. D. M. Vianna$^{a,d}$ \\
${}^{a}$Instituto deF\'{\i}sica, Universidade de Bras\'{\i}lia\\
70910-900, Bras\'{\i}lia, DF, Brasil\\
$^{b}$Physics Department, Theoretical Physics Institute,\\
University of Alberta, Edmonton, Alberta T6G 2J1 Canada\\
$^{c}$TRIUMF, 4004, Westbrook mall, Vancouver, British \\
Columbia V6T 2A3, Canada\\
$^{d}$ Instituto de F\'{\i}sica, Universidade Federal da Bahia,\\
Campus de Ondina, 40210-340, Salvador, Bahia, Brasil.}
\maketitle

\begin{abstract}
In this work we study  representations of the Poincar\'{e} group defined
over symplectic manifolds, deriving the Klein-Gordon and the Dirac equation in  
phase space. The  formalism is associated with relativistic
Wigner functions; the Noether theorem is derived in phase space
and an interacting field, including a gauge field,   approach is discussed.
\end{abstract}

\section{Introduction}
Noncommutative geometry has its origin in the
 Weyl and Moyal works,
studying  quantization procedures
in phase space \cite{moy1}.  Snyder\cite{sny1} was the first to develop
a consistent theory for non-commutative space coordinates, which was based on
representations
of
Lie Algebras.
Over the last decades
there is
a revival of
noncommutative physics, motivated by some results coming from gravity,
condensed matter physics and string theory\cite{sny3,sny4}. One particular
interest in this context
is the development of representation theories for noncommutative fields 
\cite{sny5}.
However, this type of improvement has not been fully explored  in the context
of
the Weyl-Moyal (phase space) program. In this paper we address this problem,
following
a recent work by us \cite{seb1}, where we have considered a representation for
the Galilei group on a symplectic manifold. As a result the Schr\"{o}dinger
(not
the Liouville-von Neumann ) equation was derived in phase space closely
associated with
the Wigner function. The formalism was used to treat a non-linear oscillator
perturbatively  and  to analyze the concept of coherent states from this
phase-space point of view.

 The notion of phase space in quantum mechanics arose with the paper by Wigner
 \cite{wig1} in order to develop the quantum kinetic theory.
 In the Wigner formalism, each operator, say $A$,
defined in the Hilbert space, $\mathcal{H}$, is associated with a function,
say $a_{W}(q,p)$, in phase space, $\Gamma $. Then there is an application
$\Omega
_{W}:A\rightarrow a_{W}(q,p)$, such that, the associative algebra of
operators defined in $\mathcal{H}$ turns out to be an associative (but not
commutative) algebra in $\Gamma ,$ given by $\Omega _{W}:AB\rightarrow
a_{W}\ast b_{W},$ where the star (or Moyal)-product $\ast \,$\ is defined
by\cite{wig1,wig2,wig3,wig4}
\begin{equation}
a_{W}\ast b_{W}=a_{W}(q,p)\exp \left[ \frac{i }{2}(\frac{\overleftarrow{  
\partial }}{\partial q}\frac{\overrightarrow{\partial }}{\partial p}-\frac{  
\overleftarrow{\partial }}{\partial p}\frac{\overrightarrow{\partial }}{ 
\partial q})\right] b_{W}(q,p).  \label{dessa1}
\end{equation}
(Throughout this paper we use the natural units: $\hbar =c=1 $.)
Note that Eq.(\ref{dessa1}) can be seen as an operator $\widehat{A}=a_{W}\ast $
acting on functions $b_W,$   such that $\widehat{A}(b_{W})= a_{W}\ast b_{W}.$

From a mathematical and physical standpoints, the quantum phase space and the 
Moyal
product have been explored  along different ways \cite{wig2}-\cite{bos}. 
 However, it should be of 
interest to study irreducible unitary representations of kinematical
groups considering operators of the type $ \widehat{A}.$ 
 This was our procedure in Ref.  \cite{seb1},  and here we
extend
those
representations
 to the relativistic case with
Poincar\'{e} Lie group. In other words, by using the notion of symplectic
structure and Moyal
product, we construct unitary representations for the Poincar\'e-Lie
algebra, from which we  derive the Klein-Gordon and the Dirac equations in 
phase
space. The connection of our formalism with relativistic Wigner 
functions is then presented,  some aspects 
of interating fields are discussed, including gauge fields, and the Noether
theorem
is derived in $\Gamma $.
It is worth mentioning that this alternate version for
the  field theory provides a way to consider the Wigner approach on the bases
of group representation of kinematical symmetries.

The presentation is organized in the following way. In Section 2, we define
a Hilbert space $\mathcal{H}(\Gamma )$ over a phase space $\Gamma$ with its
natural
relativisitic symplectic structure. $\mathcal{H}(\Gamma )$ will turn out to
be the space of representations of kinematical symmetry.  
In section 3 we study the Poincar\'{e} algebra in $\mathcal{H} %
(\Gamma )$ and the representations for spin zero and spin 1/2. The Noether
theorem is derived in Section 4. The developments of the
interacting symplectic field formalism is presented, for   a scalar and
gauge field,
in Section 5; and
in
Section 6 we present our final concluding remarks.

\section{Hilbert Space and Symplectic Structure}

Consider $M$ an analytical manifold where each point is specified by
Minkowski coordinates $q^{\mu },$ with $\mu = 0,1,2,3$ and  metric specified by
 $diag(g)\mathbf{%
=(-+++)}$. The coordinates of each point in the cotangent-bundle $T^{\ast }M$
will be denoted by $(q^{\mu },p_{\mu })$. The space $T^{\ast }M$ is equipped
with a symplectic structure via a 2-form 
\begin{equation}
\omega =dq^{\mu }\wedge dp_{\mu },  \label{simp1}
\end{equation}%
called the symplectic form (sum over repeated indices is assumed). We
consider the following bidifferential operator on  $C^{\infty }(T^{\ast }M)$
functions, 
\begin{equation}
\Lambda =\frac{\overleftarrow{\partial }}{\partial q^{\mu }}\frac{%
\overrightarrow{\partial }}{\partial p_{\mu }}-\frac{\overleftarrow{\partial 
}}{\partial p^{\mu }}\frac{\overrightarrow{\partial }}{\partial q_{\mu }},
\label{fasenova2}
\end{equation}%
such that for $C^{\infty }$ functions, $f(q,p)$ and $g(q,p),$ we have 
\begin{equation}
\omega (f\Lambda ,g\Lambda )=f\Lambda g=\{f,g\},  \label{fasenova3}
\end{equation}%
where 
\begin{equation*}
\{f,g\}=\frac{\partial f}{\partial q^{\mu }}\frac{\partial g}{\partial
p_{\mu }}-\frac{\partial f}{\partial p^{\mu }}\frac{\partial g}{\partial
q_{\mu }}
\end{equation*}%
is the Poisson bracket and $f\Lambda $ and $g\Lambda $ are two vector fields
given by $h\Lambda =X_{h}=-\{h,\} .$

The space $T^{\ast }M$ endowed with this symplectic structure is called the
phase space, and will be denoted by $\Gamma $.

In order to construct a Hilbert space over (this relativistic phase space) $%
\Gamma $, let $\eta $ be an invariant measure on the cotangent bundle. Then,
if $\varphi $ is a mapping: $\Gamma \rightarrow \mathbb{R}$ which is
measurable one can define the integral 
\begin{equation}
\int_{\Omega }\varphi (\mathbf{z})d\eta (\mathbf{z})
\end{equation} 
of $\varphi $ with respect to $\eta $, where $\mathbf{z}\in \Gamma $. Let $%
\mathcal{H}(\Gamma )$ be a linear subspace of the space of $\eta -measurable$
functions $\psi :\Gamma \rightarrow \mathbb{C}$ which are square integrable,
i.e. such that 
\begin{equation}
\int_{\Gamma }\mid \psi (\mathbf{z})\mid ^{2}d\eta (\mathbf{z})<\infty .
\end{equation}%
We can then introduce a Hilbert space inner product, $\langle \cdot |\cdot
\rangle $, on $\mathcal{H}(\Gamma )$, as follows:
\begin{equation}
\langle \psi _{1}|\psi _{2}\rangle =\int_{\Gamma }\psi _{1}(q,p)^{\dagger }\psi
_{2}(q,p)d\eta (q,p),
\end{equation}%
where we take $\mathbf{z}=(q^{\mu },p_{\mu })=(q,p).$ (We are using the
notation
$\psi^{\dagger}(q,p)$ for
the complex conjugation.) 

Now we take  $d\eta (q,p)=d^{4}pd^{4}q $, such that 
\begin{equation}
\int d^{4}pd^{4}q\psi^{\dagger}(q,p)\psi(q,p) < \infty,
\end{equation}
which is a real bilinear form. In this case we can write
$\psi(q,p)=\langle q,p|\psi\rangle$, with

\begin{equation}
\int d^{4}pd^{4}q |q,p\rangle\langle q,p|=1,
\end{equation}
where  the kets $|q,p\rangle$ are defined from the set 
of commuting operators  $\bar Q$ and $\bar P$ defined by \cite{seb1}

$$
\bar Q|q,p\rangle = q|q,p\rangle, \ \ \ \ \bar P|q,p\rangle = p|q,p\rangle.
$$
The state of a system will be described by functions $\phi(q,p)$, under the
condition
\begin{equation}
\langle \psi|\phi\rangle=\int
d^{4}pd^{4}q\psi^{\dagger}(q,p)\phi(q,p) < \infty 
\end{equation}
This Hilbert space is denoted by $\mathcal{H}(\Gamma)$. 

Let us see how to study the Poincar\'e group taking  
$\mathcal{H}(\Gamma)$ as the representation space. For this purpose, we
construct unitary
transformations $U:\mathcal{H}(\Gamma)\rightarrow \mathcal{H}(\Gamma)$ such
that $\langle \psi_{1}|\psi_{2}\rangle$ is invariant. Starting with $\Lambda$
defined by Eq.(\ref{fasenova2}), a mapping $e^{i \Lambda /2 }=\ast :$ \ $%
\Gamma \times \Gamma \rightarrow \Gamma ,$ called Moyal (or star) product,
is defined by 
\begin{eqnarray}
f(q,p)\ast g(q,p) &=&f(q,p)\exp \left[ \frac{i  }{2}\left(
\overleftarrow{\partial 
}_{q}\overrightarrow{\partial }_{p}-\overleftarrow{\partial }_{p}%
\overrightarrow{\partial }_{q}\right) \right] g(q,p)  \label{A98} \\
&=&\exp \left[ \frac{i  }{2}\left( \partial _{q}\partial _{p^{\prime
}}-\partial _{p}\partial _{q^{\prime }}\right) \right] f(q,p)g(q^{\prime
},p^{\prime })|_{q^{\prime },p^{\prime }=q,p},  \label{a99}
\end{eqnarray}
where $f(q,p)$ and $g(q,p)$ are in $\Gamma $ and $\partial _{x}=\partial
/\partial x$ $%
(x=p,q).$  

The generators of $U$ can be introduced by the following (Moyal-Weyl)
star-operators:
\begin{equation}
\widehat{F} = f(q,p)*=f(q^{\mu}+\frac{i}{2}\frac{\partial}{\partial
p_{\mu}}, p^{\mu}-\frac{i}{2}\frac{\partial}{\partial
q_{\mu}} ) \label{star11}.
\end{equation}
In the sequence we analyze the Poincar\'e   algebra in this context.

\section{Poincar\'e-Lie algebra in $\mathcal{H}(\Gamma)$}
Using the functions  $q_{\mu}$, $p_{\mu}$, and $m_{\mu \nu}=q_{\mu} p_{\nu}
-p_{\mu}q_{\nu}$ (All of them defined in $\Gamma$), and equation
(\ref{star11}),
we construct the corresponding star operators
\begin{equation}
\widehat{P}^{\mu}=p^{\mu}\ast = p^{\mu}-\frac{i}{2}\frac{\partial}{\partial
q_{\mu}}, \label{p}
\end{equation}

\begin{equation}
\widehat{Q}^{\mu}=q^{\mu}\ast = q^{\mu}+\frac{i}{2}\frac{\partial}{\partial
p_{\mu}},
\end{equation}
and
\begin{equation}
\widehat{M}_{\nu\sigma}=M_{\nu\sigma}*
=\widehat{Q}_{\nu}\widehat{P}_{\sigma}-\widehat{Q}_{\sigma}\widehat{P}_{\nu}.
\end{equation}
From this set of unitary operators we obtain, after some long but simple
calculations, the following set of commutation relations,
\begin{equation}\nonumber
[\widehat{M}_{\mu\nu},\widehat{P}_{\sigma}]=i(g_{\nu\sigma}\widehat{P}_{\mu}-g_{
\sigma\mu}\widehat{P}_{\nu}),
\end{equation}

\begin{equation}\nonumber
[\widehat{P}_{\mu},\widehat{P}_{\nu}]=0,
\end{equation}

\begin{equation}\nonumber
[\widehat{M}_{\mu\nu},\widehat{M}_{\sigma\rho}]=-i(g_{\mu\rho}\widehat{M}_{\nu\sigma}-
g_{\nu\rho}\widehat{M}_{\mu\sigma}+g_{\mu\sigma}\widehat{M}_{\rho\nu}-g_{\nu 
\sigma}\widehat{M}_{\rho\mu}).
\end{equation}
This is  the Poincar\'e  algebra, where $\widehat{M}_{\mu\nu}$ stand for
rotations and 
$\widehat{P}_{\mu}$ for translations (but notice, in phase space).

The Casimir invariants can be built up from the Pauli-Lubanski matrices,  
$\widehat{W}_{\mu }=\frac{1}{2}\varepsilon _{\mu \nu \rho \sigma
}\widehat{M}^{\nu \sigma 
}\widehat{P}^{\rho }$, where $\varepsilon _{\mu \nu \rho \sigma }$ is the
Levi-Civita
symbol. The invariants are: 
\begin{eqnarray}
\widehat{P}^{2} &=&\widehat{P}^{\mu }\widehat{P}_{\mu }, \label{inv4} \\
\widehat{W} &=&\widehat{W}^{\mu }\widehat{W}_{\mu },  \label{inv1}   
\end{eqnarray}
where $\widehat{P}^{2} $ stands for the mass shell condition and $\widehat{W}$
for the spin. In the following we use such a representation to derive equations
in phase space for spin 0 and spin 1/2 particles.

To determine the Klein-Gordon field equation, we consider a scalar 
representation
in $\mathcal{H}(\Gamma)$. In this case we can use the invariant $ \widehat{P}
^2
$ given in Eq.(\ref{inv4}) to write down 
\begin{eqnarray*}
\widehat{P}^2\psi(q,p)&=&(p^2)\ast \phi(q,p) \\
&=&(p^{\mu }\ast p_{\mu }\ast) \phi(q,p)=m^2\phi(q,p),
\end{eqnarray*}
where $m$ is a constant fixing the representation and interpreted as mass,
such that the mass shell condition is satisfied. Using Eq.(\ref{p}) we obtain 
\begin{equation}
(p^{\mu }p_{\mu }-ip^{\mu}\frac{\partial }{\partial q^{\mu }}-\frac{1}{4}%
\frac{\partial }{\partial q^{\mu }}\frac{\partial }{\partial q_{\mu }}%
)\phi(q,p)=m^2\phi(q,p) ,  \label{kgeq1}
\end{equation}
which is the Klein-Gordon equation in phase space. The solution for such
equation is 
\begin{equation}
\phi(q_{\mu}, p_{\mu})=\xi(p_{\mu})e^{-i4p^{\mu}q_{\mu}},
\end{equation}
where $\xi(p_{\mu})$ is a function that depends on the boundary-conditions.

The  Lagrangian that leads to Eq.(\ref{kgeq1}) is given by
\begin{equation}
\pounds = \frac{-1}{4}\frac{\partial\phi}{\partial
q_{\mu}}\frac{\partial\phi^{\dagger}}{\partial q^{\mu}} +
\frac{1}{2}ip^{\mu}(\phi^{\dagger}\frac{\partial\phi}{\partial
q^{\mu}}-\phi \frac{\partial\phi^{\dagger}}{\partial q^{\mu}}) -
(p^{\mu}p_{\mu}-m^{2})\phi\phi^{\dagger}.\label{kgeq111}
\end{equation}
We  use later this Lagrangian to analyze  the Noether theorem in phase space
and the interaction  with gauge fields.

The association of this representation with the  Wigner formalism is given by
\begin{equation*}
f_{W}(q,p)=\phi(q,p)\ast\phi^{\dagger}(q,p),  
\end{equation*}
where $f_{W}(q,p)$ is the relativistic Wigner function. To prove this, we
recall
 that the Klein-Gordon equation
in   phase space can be written as 
\begin{equation}
\widehat{P}^{2} \phi=p^{2}\ast\phi=m^{2}\phi .  \label{KGE}
\end{equation}
Multiplying the right hand side of the above equation by $\phi^{\dagger}$ we
obtain 
\begin{equation}
(p^{2}\ast\phi)\ast\phi^{\dagger}=m^{2}\phi\ast\phi^{\dagger} ,
\label{KGEMR}
\end{equation}
but since $\phi^{\dagger}\ast p^{2}=m^2\phi^{\dagger}$, we also have 
\begin{equation}
\phi\ast(\phi^{\dagger}\ast p^{2})=m^2\phi \ast \phi^{\dagger}.
\label{KGEML}
\end{equation}
Subtracting (\ref{KGEMR}) of (\ref{KGEML}), and using the associativity
property
of
the ``$\ast$" product, we get 
\begin{equation}
p^{2}\ast f_{W}-f_{W}\ast p^{2}=0,  \label{cpw}
\end{equation}
where the notation $f_{W}=\phi \ast \phi^{\dagger}$ has been used.

Equation (\ref{cpw}) can be written as the relativistic Moyal bracket, i.e.
\begin{equation*}
\lbrace g,f \rbrace_{M}=(q,p)\ast f(q,p)- f(q,p)\ast g(q,p)=g\left(2\sin
\frac{i}{2}\Lambda \right)f.
\end{equation*}
Applying this results to Eq.(\ref{cpw}), we obtain 
\begin{equation}
 p_{\mu}\frac{\partial}{\partial q_{\mu}}f_{W}=0 ,  \label{flux1}
\end{equation}
a well known result. Other properties of the Wigner function, such as the
non-positiveness, can be derived as in the non-relativistic  case; see
Ref.\cite{seb1}.

Now let us turn our attention to spin 1/2 particles. Proceeding as usual, we
assume the invariant operator 
$\gamma^{\mu}\widehat{P} _{\mu}$,  where $\widehat{P}_{\mu}$ is defined by
Eq.(\ref{p}).
Thus we write
\begin{equation}\nonumber
\gamma^{\mu}\widehat{P}_{\mu}\psi=\gamma^{\mu}p_{\mu}\ast\psi=m\psi.
\end{equation}
Using Eq.(\ref{p}), we obtain
\begin{equation}\label{dir11}
\gamma^{\mu}(p_{\mu}-\frac{i}{2}\frac{\partial}{\partial
q^{\mu}})\psi=m\psi,
\end{equation}
which is the Dirac equation in phase space, where the
$\gamma^{\mu}$-matrices fullfill the usual Clifford algebra,
$
(\gamma^{\mu}\gamma^{\nu}+\gamma^{\nu}\gamma^{\mu})=2g^{\mu\nu}.
$

The Lagrangian for such a field can be written as 
\begin{equation}
\pounds=\frac{-i}{4}(\frac{\partial \overline{\psi}}{\partial
q^{\mu}}\gamma^{\mu}\psi -
\overline{\psi}\gamma^{\mu}\frac{\partial\psi}{\partial q^{\mu}   
})-\overline{\psi}(m-\gamma^{\mu}p_{\mu}) {\psi}.
\end{equation}
where $ \overline{\psi}= \psi^{\dagger}(q,p) \gamma_0$, with 
$\psi^{\dagger}(q,p)$ being the Hermitian adjoint of $\psi(q,p).$
The Wigner
function
in
this case is given by
$f_W(q,p)=\psi(q,p)*\overline{\psi}(q,p)$, with each component satisfying
Eq.(\ref{flux1}).
%
%

\section{Noether Theorem in Phase Space}
Our proposal in this section is to demonstrate the Noether theorem in phase
space, stating the following, 
\newline

\textbf{Noether theorem in $\mathcal{H}(\Gamma)$}: {\textit {To all the
Lie-transformation group changing $\Psi (q,p)$ but leaving $\pounds (\Psi, 
 \partial{\Psi}    )$
invariant
up to a divergence, there exists a conserved current in phase space. The
divergence is defined by}}

$$ (\frac{\partial}{\partial q^{\mu}}+ \frac{\partial}{\partial
p^{\mu}})S^{\mu}. $$
Here $\Psi$ stands for the Klein-Gordon field, $\phi,$ or the
Dirac field, $\psi.$

\textbf{Proof} : Consider the infinitesimal transformation
$\Psi(q,p)\rightarrow
\Psi'(q,p)=\Psi(q,p)+\delta\Psi(q,p)$, such that $\pounds\rightarrow
\pounds+\delta\pounds$, where  $\delta\pounds=(\frac{\partial
}{\partial q^{\mu}}+ \frac{\partial }{\partial p^{\mu}})S^{\mu}.$
Then,
$$\delta\pounds=\frac{\partial\pounds}{\partial\Psi}\delta\Psi+
\frac{\partial\pounds}{\partial(\frac{\partial\Psi}{\partial
q^{\mu}})}\frac{\partial\delta\Psi}{\partial
q^{\mu}}+\frac{\partial\pounds}{\partial(\frac{\partial\Psi}{\partial
p^{\mu}})}\frac{\partial\delta\Psi}{\partial
p^{\mu}}.
$$
Developing this equation and using the Euler-Lagrange equation of motion in
phase space i.e, 
$$\frac{\partial\pounds}{\partial\Psi}-\frac{\partial}{\partial
q^{\mu}}\frac{\partial\pounds}{\partial(\frac{\partial\Psi}{\partial
q^{\mu}})}-\frac{\partial}{\partial
p^{\mu}}\frac{\partial\pounds}{\partial(\frac{\partial\Psi}{\partial
p^{\mu}})}=0,$$ 
we get 
\begin{eqnarray}
\delta\pounds&=&(\frac{\partial}{\partial
q^{\mu}}\frac{\partial\pounds}{\partial(\frac{\partial\Psi}{\partial
q^{\mu}})}\delta\Psi+\frac{\partial}{\partial
p^{\mu}}\frac{\partial\pounds}{\partial(\frac{\partial\Psi}{\partial
p^{\mu}})}\delta\Psi)+
(\frac{\partial\pounds}{\partial(\frac{\partial\Psi}{\partial
q^{\mu}})}\frac{\partial\delta\Psi}{\partial
q^{\mu}}+\frac{\partial\pounds}{\partial(\frac{\partial\Psi}{\partial
p^{\mu}})}\frac{\partial\delta\Psi}{\partial p^{\mu}})\nonumber
\\&=&\frac{\partial}{\partial
q^{\mu}}(\frac{\partial\pounds}{\partial(\frac{\partial\Psi}{\partial
q^{\mu}})}\delta\Psi)+ \frac{\partial}{\partial
p^{\mu}}(\frac{\partial\pounds}{\partial(\frac{\partial\Psi}{\partial
p^{\mu}})}\delta\Psi).\nonumber
\end{eqnarray}
As a consequence we obtain
$$(\frac{\partial}{\partial q^{\mu}}+ \frac{\partial}{\partial
p^{\mu}})j^{\mu}=0,$$
where the conserved current is
\begin{equation}\nonumber
j^{\mu}=
\frac{\partial\pounds}{\partial(\frac{\partial\Psi}{\partial
q^{\mu}})}\delta\Psi
+\frac{\partial\pounds}{\partial(\frac{\partial\Psi}{\partial
p^{\mu}})}\delta\Psi -S^{\mu}.  
\end{equation}
This provides a proof of the existence of the Noether-current in phase space.
Let us use this result to study some particular transformations.
\newline

\textbf{Example (i): $\Gamma$-time  translations}. Consider
$$q^{\mu}\rightarrow q'^{\mu}=q^{\mu}+\varepsilon^{\mu}, 
 p^{\mu}\rightarrow p'^{\mu}=p^{\mu}+\lambda^{\mu}.$$
with the notation
$\eta^{\mu}=(q^{\mu},p^{\mu})$, as well as
$\epsilon^{\mu}=(\varepsilon^{\mu},\lambda^{\mu})$.
Then we obtain 
$$
(\frac{\partial}{\partial q^{\mu}}+ \frac{\partial}{\partial
p^{\mu}})\theta^{\mu\nu}=0,
$$ 
where the  $\Gamma$-stress tensor, $\theta^{\mu\nu}$, is given by
\begin{equation*} 
\theta^{\mu\nu}=\frac{\partial\pounds}{\partial(\frac{\partial\Psi}{\partial
\eta^{\mu}})}\frac{\partial\Psi}{\partial
\eta^{\nu}}-g^{\mu\nu}\pounds.
\end{equation*}

Taking, as an example, space translations only, that is,
$\lambda=0$,  we obtain for the free Klein-Gordon field
\begin{eqnarray*}
\theta^{\mu\nu}_{KG}&=&\frac{-1}{4}(\frac{\partial\phi^{\ast}}{\partial
q_{\mu}}\frac{\partial\phi}{\partial
q_{\nu}}+\frac{\partial\phi}{\partial
q_{\mu}}\frac{\partial\phi^{\ast}}{\partial
q_{\nu}})+\frac{1}{2}ip^{\mu}(\phi^{\ast}\frac{\partial\phi}{\partial
q_{\nu}}-\phi\frac{\partial\phi^{\ast}}{\partial
q_{\nu}})-g^{\mu\nu}\pounds.
\end{eqnarray*}
For the Dirac field we obtain
\begin{equation}\nonumber
\theta^{\mu\nu}_{D}=\frac{-i}{4}(-\overline{\psi}\gamma^{\mu}\frac{\partial\psi}
{\partial
q_{\nu}} + \gamma^{\mu}\psi\frac{\overline{\psi}}{\partial
q_{\nu}})-g^{\mu\nu}\pounds. \label{noet1}
\end{equation}
In these two examples, we obtain the usual invariant 4-momentum
$
P^{\nu}=\int \theta^{0 \nu}d^{3}q d^{4}p,
$
\newline

\textbf{Example (ii): $\Gamma$-time  rotations.} 
We define   rotations in $\Gamma $ by $q^{\nu}\rightarrow q^{\nu}+
\epsilon^{\mu\nu}q_{\nu},$ and  $p^{\nu}\rightarrow p^{\nu}+
\lambda^{\mu\nu}p_{\nu},$ such that 
$\delta q^{\mu}=\epsilon^{\mu\nu}q_{\nu},$ and 
$\delta p^{\mu}=\lambda^{\mu\nu}p_{\nu},$
with 
$\epsilon^{\mu\nu}=-\epsilon^{\nu\mu},$
and
$\lambda^{\mu\nu}=-\lambda^{\nu\mu}.$
Then we find
\begin{equation}\nonumber
(\frac{\partial}{\partial q^{\mu}}+ \frac{\partial}{\partial
p^{\mu}})M^{\mu\nu\lambda}=0,
\end{equation}
where $M^{\mu\nu\lambda}$ is a $\Gamma$-angular-momentum tensor given by
\begin{equation}\nonumber
M^{\mu\nu\lambda}=-q^{\lambda}\theta^{\mu\nu}-p^{\lambda}\theta^{\mu\nu}+q^{\nu}
\theta^{\mu\lambda}+p^{\nu}\theta^{\mu\lambda}+i\frac{\partial\pounds}{\partial(
\frac{\partial\Psi}{\partial
q^{\mu}})}I^{\nu\lambda}\Psi+
i\frac{\partial\pounds}{\partial(\frac{\partial\Psi}{\partial
p^{\mu}})}I^{\nu\lambda}\Psi,
\end{equation}
 with $I^{\mu\nu}$ being such that
$$
\delta \Psi=-\frac{\epsilon^{\mu \nu}}{2}(
\eta_{\nu}\partial_{\mu}-\eta_{\mu}\partial_{\nu} -iI_{\mu \nu}
)\Psi(\eta).
$$
For the Klein-Gordon and Dirac fields, respectively, we thus have 

\begin{eqnarray*}
M^{\mu\nu\lambda}_{KG}&=&-q^{\lambda}\theta_{KG}^{\mu\nu}-p^{\lambda}\theta_{KG}
^{\mu\nu}
+q^{\nu}\theta_{KG}^{\mu\lambda}+p^{\nu}\theta_{KG}^{\mu\lambda} + 
i\frac{\partial\phi^{\ast}}{\partial
q_{\mu}}I^{\nu\lambda}\phi\\
&&+i\frac{\partial\phi}{\partial
q_{\mu}}I^{\nu\lambda}\phi^{\ast}
 -\frac{1}{2}p^{\mu}\phi^{\ast}I^{\nu\lambda}\phi+\frac{1}{2}p^{\mu}\phi
I^{\nu\lambda}\phi^{\ast},
\end{eqnarray*}
and
\begin{equation}\nonumber
M^{\mu\nu\lambda}_{D}=-q^{\lambda}\theta_{D}^{\mu\nu}-p^{\lambda}\theta_{D}^{\mu
\nu}+q^{\nu}\theta_{D}^{\mu\lambda}+p^{\nu}\theta_{D}^{\mu\lambda}-\frac{1}{4}
\overline{\psi}\gamma^{\mu}I^{\nu\lambda}\psi+\frac{1}{4}\gamma^{\mu}\psi
I^{\nu\lambda} \overline{\psi},
\end{equation}

The usual (space rotations only) angular momentum is then given by
$
\emph{M}^{\mu\nu}=\int M^{\mu\nu 0} d^{3}q d^{4}p.
$
\newline

\textbf{Example (iii): non-local gauge symmetry}. Using the Noether theorem,
let
us start with the analysis of gauge symmetries in the context of this
symplectic
field theory. We consider a global gauge transformation given by
$$
\Psi'=e^{-i\alpha}\Psi   \quad    and
\quad\overline{\Psi'}=e^{i\alpha}\overline{\Psi},$$
where $\alpha$
is a real constant. Up to first order, we have
$$\Psi'=(1-i\alpha)\Psi \quad \ \ \  and \ \ \ \quad
\delta\Psi=-i\alpha\Psi.$$ we then assume
$\delta\pounds=0$. Then for the Klein-Gordon and Dirac field, respectively, we
have
\begin{equation}
j_{KG}^{\mu}=\frac{i}{4}(\phi\frac{\partial\phi^{\ast}}{\partial
q_{\mu}}+\phi^{\ast}\frac{\partial\phi}{\partial q_{\mu}}) +
p^{\mu}\phi^{\ast}\phi.
\end{equation}
and 
\begin{equation}
j_{D}^{\mu}=\overline{\psi}\gamma^{\mu}\psi.
\end{equation}
 For these two cases,  we have
$
Q=\int j^{0} d^{3}q d^{4}p.
$
%
%
%
\section{Elements of Interacting Fields}
 In this section we consider the problem of interacting
 fields in phase
space. Let us
consider the scalar-field  Lagrangian given by
\begin{equation}\nonumber
\pounds = \frac{-1}{4}\frac{\partial\phi}{\partial
q_{\mu}}\frac{\partial\phi^{\dagger}}{\partial q^{\mu}} +
\frac{1}{2}ip^{\mu}(\phi^{\dagger}\frac{\partial\phi}{\partial
q^{\mu}}-\phi \frac{\partial\phi^{\dagger}}{\partial q^{\mu}}) -
(p^{\mu}p_{\mu}-m^{2})\phi\phi^{\dagger}+ U(\phi \phi^{\dagger}),
\end{equation}
that leads to the following non-linear Klein-Gordon equation,
\begin{equation}\nonumber
\frac{-1}{4}\frac{\partial^{2}\phi}{\partial q^{\mu}\partial
q_{\mu}}-ip^{\mu}\frac{\partial\phi}{\partial
q^{\mu}}+(p^{\mu}p_{\mu}-m^{2})\phi+V(\phi)=0,  
\end{equation}
where $V(\phi)=\frac{\partial
U(\phi \phi^{\dagger})}{\partial\phi^{\dagger}}$.
In order to solve this equation, we can use the Green's function
$G=G(q^{\mu},q'^{\mu},p^{\mu},p'^{\mu})$ satisfying the equation
\begin{equation}\nonumber
\frac{-1}{4}\frac{\partial^{2}G}{\partial q^{\mu}\partial
q_{\mu}}-ip^{\mu}\frac{\partial G}{\partial
q^{\mu}}+(p^{\mu}p_{\mu}-m^{2})G=\delta(q^{\mu}-q'^{\mu})\delta(p^{\mu}-p'^{\mu}
),
\end{equation}
Then a solution of such equation is given by
\begin{equation}\nonumber
\phi(q^{\mu},p^{\mu})= \phi_{0}(q^{\mu},p^{\mu})+\int
d^{4}q'^{\mu}d^{4}p'^{\mu}G(q^{\mu},q'^{\mu},p^{\mu},p'^{\mu})V(\phi),
\end{equation}
where $\phi_{0}(q^{\mu},p^{\mu})$ is the solution of the homogeneous
Klein-Gordon equation in phase space. The poles of the Green's function can be 
derived by the
following. Taking a Fourier transform, such that
$q^{\mu}\rightarrow k^{\mu}$ and $p^{\mu}\rightarrow \xi^{\mu}$,  we obtain,
\begin{equation}\nonumber
\frac{1}{4}k^{2}\widetilde{G}(k^{\mu},\xi^{\mu}) -
p^{\mu}k_{\mu}\widetilde{G}(k^{\mu},\xi^{\mu})+
(p^{\mu}p_{\mu}-m^{2})\widetilde{G}(k^{\mu},\xi^{\mu})=1,
\end{equation}
where
\begin{equation}\nonumber
\widetilde{G}(k^{\mu},\xi^{\mu})=\frac{1}{\frac{1}{4}k^{2} -
p^{\mu}k_{\mu}+ p^{\mu}p_{\mu}-m^{2}}.
\end{equation}
Therefore the Green's function is written as
\begin{equation}\nonumber
G(q^{\mu},q'^{\mu},p^{\mu},p'^{\mu})=\frac{1}{(2\pi)^{4}}\int
d^{4}k^{\mu}d^{4}\xi^{\mu}\frac{e^{-ik^{\mu}q_{\mu}-i\xi^{\mu}p_{\mu}}}{\frac{1}
{4}k^{2}
- p^{\mu}k_{\mu}+  p^{\mu}p_{\mu}-m^{2} }.
\end{equation}

With this result,  we construct a generating
functional, starting then a study of a quantum field theory in phase space.
First, we define  the Feynman propagator by
$$
(\square +ip\partial+p^2+m^2+i\varepsilon )G^F_{0}(q,p)=-\delta (q,p),
$$

where $ip\partial=ip^{\mu}/\partial q^{\mu}.$
The  generator functional is introduce by
\begin{eqnarray*}
Z_{0} &\simeq &\int D\phi D\widetilde{\phi }e^{iS}=
\int D\phi  \exp [i\int dqdp(\pounds)] \\
&=&\int D\phi  \exp \{-i\int dqdp[\frac{1}{2}\phi(q,p) (\square
+ip\partial+p^2+m^2+i\varepsilon)\phi(q,p) -J(q,p)\phi(q,p)
\}.
\end{eqnarray*}%
Such a functional can  be written as
\begin{equation}
Z_{0}\simeq \exp \{\frac{i}{2}\int dqdpdq'dp'[J(q,p)(\square
+ip\partial+p^2+m^2+i\varepsilon
)^{-1}J(q',p')\}.  \label{andr1}
\end{equation}%
Then we find
\begin{equation*}
G^F_{0}(\eta-\eta';\beta )=i\frac{\delta
^{2}Z[J]}{\delta J (\eta)\delta J(\eta')}|_{J=0},
\end{equation*}
where again we have used the  notation, $\eta=(x,p))$.

Another important aspect of the interacting fields, that can be analyzed in 
phase space
with
standard  methods, is the gauge symmetry. We start with the
 scalar
field considering a local phase transformation given by
$$\phi'(q,p)=e^{-i\alpha(q,p)}\phi(q,p).$$
Taking an infinitesimal $\alpha$, we have
$\phi\rightarrow \phi-i\alpha\phi, \ \
  \delta\phi=-i\alpha\phi, $ $ \ \ \delta\phi^{\dagger}=i\alpha\phi^{\dagger},$
such that
$$\delta[(\frac{\partial}{\partial
q^{\mu}}+\frac{\partial}{\partial
p^{\mu}})\phi]=-i[(\frac{\partial}{\partial
q^{\mu}}+\frac{\partial}{\partial
p^{\mu}})\alpha]\phi-i\alpha(\frac{\partial}{\partial
q^{\mu}}+\frac{\partial}{\partial p^{\mu}})\phi,$$
and
$$\delta[(\frac{\partial}{\partial
q^{\mu}}+\frac{\partial}{\partial
p^{\mu}})\phi^{\dagger}]=i[(\frac{\partial}{\partial
q^{\mu}}+\frac{\partial}{\partial
p^{\mu}})\alpha]\phi^{\dagger}+
i\alpha(\frac{\partial}{\partial
q^{\mu}}+\frac{\partial}{\partial p^{\mu}})\phi^{\dagger}.$$
Using the Lagrangian given in Eq.(\ref{kgeq111})
for the free scalar field, we obtain
$$\delta\pounds=i[\frac{i}{4}(\phi\frac{\partial\phi^{\dagger}}{\partial
q_{\mu}}+\phi^{\dagger}\frac{\partial\phi}{\partial q_{\mu}}) +
p^{\mu}\phi^{\dagger}\phi]\partial_{\mu}\alpha=j^{\mu}\partial_{\mu}\alpha,$$

Therefore, demanding invariance of the Lagrangian under the gauge
transformation, we introduce
the $U(1)$ gauge field, such  that the full Lagrangian
 reads
\begin{eqnarray}
\pounds_{total}&=&\frac{-1}{4}\frac{\partial\phi}{\partial
q_{\mu}}\frac{\partial\phi^{\dagger}}{\partial q^{\mu}} +
\frac{1}{2}ip^{\mu}(\phi^{\dagger}\frac{\partial\phi}{\partial
q^{\mu}}-\phi \frac{\partial\phi^{\dagger}}{\partial q^{\mu}}) -
(p^{\mu}p_{\mu}-m^{2})\phi\phi^{\dagger}\nonumber
\\&&-
ej^{\mu}A_{\mu}+
e^{2}A_{\mu}A^{\mu}\phi^{\dagger}\psi-\frac{1}{4}F^{\mu\nu}F_{\mu\nu} \\
&=&
(p_{\mu}\psi-\frac{i}{2}\frac{\partial\phi}{\partial
q^{\mu}}+ieA_{\mu}\phi)(p^{\mu}\phi^{\dagger}+\frac{i}{2}
\frac{\partial\phi^{\dagger}}{\partial
q_{\mu}}-ieA^{\mu}\psi^{\dagger})+m^{2}\phi^{\dagger}\phi-
\frac{1}{4}F^{\mu\nu}F_{\mu\nu}
,\nonumber
\end{eqnarray}
where $F_{\mu\nu}=
\partial A_{\nu}/\partial {q^{\mu}} - \partial
A_{\mu}/\partial{q^{\nu}}.$

A  covariant derivative can be defined by
$$D_{\mu}=(p_{\mu}-\frac{i}{2}\frac{\partial}{\partial
q^{\mu}}+ieA_{\mu}),$$ leading  to
$$\pounds=D_{\mu}\phi
D^{\mu}\phi^{\dagger}+m^{2}\phi\phi^{\dagger}-\frac{1}{4}F^{\mu\nu}F_{\mu\nu}.$$
It is important to notice that the minimum coupling is given by
$\widehat{P}_{\mu}\rightarrow
\widehat{P}_{\mu}+ieA_{\mu}$, with 
$\widehat{P}_{\mu}=p_{\mu}-\frac{i}{2}\frac{\partial}{\partial
q^{\mu}},$ as it is
expected. In closing, with the definition of a Feynman propagator, consistent 
perturbation calculations in phase space can be carried out; this is one aspect 
  to be considered  in another place.

\section{Concluding Remarks}
In this paper we have set forth a field theory based on a relativistic  Hilbert
phase space, using as a
basic ingredient the Moyal product of the non-commutative geometry.  We
develop
 a representation theory for kinematical Lie groups
as it was stated by us in  \cite{seb1},
considering the Galilei group; and in this sense our
procedure generalizes the   Curtright  and Zacos approach 
\cite{wig55}. As a
consequence we have derived from the Poincar\'e Lie algebra in phase space:
the Klein-Gordon equation (see Eq. (\ref{kgeq1})),  the Dirac equation
(see Eq. (\ref{dir11})),
  the Noether theorem, and  the association of these results with
the Wigner function formalism.

Furthermore, we extend the formalism to treat interacting fields.  First
  a $\lambda \phi ^4(q,p) $ theory is analyzed: we define the proper propagator
for use in perturbative
methods  and introduce a quantization scheme, delineated by the
definition of a generating functional.  Second, we consider
the abelian, $U(1),$  symmetry,  suggesting a generalization
to  non-abelian gauge fields. One central point to be emphasized is that the
approach
developed here permits the calculation of Wigner functions for relativistic
systems with methods, based on symmetry, similar to those used in quantum field
theory, including prescriptions for diagrammatic  analysis. A more detailed 
account of this
will  be considered in a longer paper.
$$$$

\textbf{Acknowledgments:}   This work was partially supported by CNPq of
Brazil and NSERC of Canada.

\end{document}